\documentstyle[12pt]{article}
\begin{document}
\large

\newpage
\begin{center}
{\bf ON THE ELECTRIC CHARGE OF THE NEUTRINO}
\end{center}
\vspace{1cm}
\begin{center}
{\bf Rasulkhozha S. Sharafiddinov}
\end{center}
\vspace{1cm}
\begin{center}
{\bf Institute of Nuclear Physics, Uzbekistan Academy of Sciences,
Tashkent, 702132 Ulugbek, Uzbekistan}
\end{center}
\vspace{1cm}

Exact expression is obtained for the differential cross section of elastic
electroweak scattering of longitudinal polarized massive Dirac neutrinos
with the electric charge and anomalous magnetic moment on a spinless
nucleus. This formula contains all necessary information about the nature
of the neutrino mass, charge and magnetic moment. Some of them state
that between the mass of the neutrino its electric charge there exists
interconnection.

\newpage
From time to time in leterature the considerations in favor of the existence
of a small electric charge $e_{\nu},$ magnetic moment $\mu_{\nu}$ or
electromagnetic radius $r_{\nu}$ of the neutrino with a non - zero rest
mass $m_{\nu}$ are listed and some possible consequences of their interaction
with other particles \cite{1,2,3,4} are considered. Reactions with electrons
and nucleons, and also the neutron and muon decays apply to them.

Exactly the same one can as the source of the information choose the
processes on the nuclear targets. In the case of a nucleus with the electric
$(Z)$ and weak $(Z_{W})$ charges on the elastic scattering \cite{5}
\begin{equation}
\nu(\overline{\nu})+A(Z, Z_{W})\stackrel{\gamma,Z^{0}}
{\rightarrow}\nu'(\overline{\nu'})+ A(Z, Z_{W}),
\label{1}
\end{equation}
influence only the properties of the neutrino itself. In this connection
appears of principle possibility to rewatch the behavior of the neutrino
in the elastic scattering on a spinless nucleus. Such an analysis is, in
particular, carried and out in the present work. It is assumed that the
neutrino has the longitudinal polarization.

The scattering amplitude at the account of hadronic and leptonic weak neutral
currents may to the lower order in $\alpha$ and $G_{F}$ be written as
$$M_{fi}^{ew}=M_{fi}^{em}+M_{fi}^{we}=$$
$$=\frac{G_{F}}{\sqrt{2}}\overline{u}(p',s')\gamma_{\mu}(g_{V_{\nu}}+
\gamma_{5}g_{A_{\nu}})u(p,s)J_{\mu}^{Z^{0}}(q)+$$
\begin{equation}
+\frac{4\pi\alpha}{q^{2}}\overline{u}(p',s')[\gamma_{\mu}F_{1\nu}(q^{2})-
\sigma_{\mu\lambda}q_{\lambda}F_{2\nu}(q^{2})]
u(p,s)J_{\mu}^{\gamma}(q),
\label{2}
\end{equation}
where $p(p')$ and $s(s')$ denote the four - momentum and helicity
of initial (final) neutrinos, $q=p-p',$ $\gamma_{\mu}$ and
$\sigma_{\mu\lambda}=(\gamma_{\mu}\gamma_{\lambda}-
\gamma_{\lambda}\gamma_{\mu}/2$ are the Dirac matrices, $g_{V_{\nu}}$ and
$g_{A_{\nu}}$ characterize the vector and axial vector coupling constants,
$J_{\mu}^{x}(q)$ imply the nucleus electromagnetic $(x=\gamma)$ and weak
neutral $(x=Z^{0})$ currents, $F_{1\nu}(q^{2})$ and $F_{2\nu}(q^{2})$
describe the neutrino Dirac and Pauli form factors which are in the static
limit $(q^{2}=0)$ reduced to the values
$$F_{1\nu}(0)=e_{\nu}, \, \, \, \, F_{2\nu}(0)=\mu_{\nu}.$$

In the case of a zero - spin nucleus, the differential cross section
of the process (\ref{1}) on the basis of (\ref{2}) one can present as
\begin{equation}
d\sigma_{ew}=d\sigma_{em}+d\sigma_{int}+d\sigma_{we}.
\label{3}
\end{equation}

First term here answers to purely electromagnetic scattering
and equal to
$$\frac{d\sigma_{em}}{d\Omega}=
\frac{1}{2}\sigma^{\nu}_{o}(1-\eta^{2}_{\nu})^{-1}\{(1+ss')F_{1\nu}^{2}+$$
\begin{equation}
+\eta^{2}_{\nu}(1-ss')[F_{1\nu}\mp
2m_{\nu}(1-\eta^{-2}_{\nu})F_{2\nu}]^{2}
tg^{2}\frac{\theta}{2}\}F_{E}^{2}(q^{2}).
\label{4}
\end{equation}

The contribution explained by the interference of neutral and electromagnetic
currents has the form
$$\frac{d\sigma_{int}}{d\Omega}=
\frac{1}{2}\rho\sigma^{\nu}_{o}(1-\eta^{2}_{\nu})^{-1}g_{V_{\nu}}\{(1+ss')
\left(1\mp s\frac{g_{A_{\nu}}}{g_{V_{\nu}}}\sqrt{1-\eta^{2}_{\nu}}\right)
F_{1\nu}+$$
\begin{equation}
+\eta^{2}_{\nu}
(1-ss')[F_{1\nu}\mp
2m_{\nu}(1-\eta^{-2}_{\nu})F_{2\nu}]
tg^{2}\frac{\theta}{2}\}F_{EV}(q^{2}).
\label{5}
\end{equation}

One can write also the cross section of purely weak transition
$$\frac{d\sigma_{we}}{d\Omega}=
\frac{E_{\nu}^{2}G_{F}^{2}}{16\pi^{2}}\{(g_{V_{\nu}}^{2}+
g_{A_{\nu}}^{2})(1+ss')cos^{2}\frac{\theta}{2}+$$
$$+\eta^{2}_{\nu}[g_{V_{\nu}}^{2}(1-ss')sin^{2}\frac{\theta}{2}-
g_{A_{\nu}}^{2}(1+ss')cos^{2}\frac{\theta}{2}]\mp$$
\begin{equation}
\mp 2sg_{V_{\nu}}g_{A_{\nu}}(1+ss')\sqrt{1-\eta_{\nu}^{2}}
cos^{2}\frac{\theta}{2}\}F_{W}^{2}(q^{2}),
\label{6}
\end{equation}
where the upper (lower) sign corresponds to the neutrino (antineutrino).
Here one must have in view of that
$$\sigma_{o}^{\nu}=
\frac{\alpha^{2}cos^{2}\frac{\theta}{2}}{4E_{\nu}^{2}(1-\eta^{2}_{\nu})
sin^{4}\frac{\theta}{2}}, \, \, \, \,
\eta_{\nu}=\frac{m_{\nu}}{E_{\nu}}, \, \, \, \,
\rho=\frac{G_{F}q^{2}}{2\pi\sqrt{2}\alpha},$$
$$F_{E}(q^{2})=ZF_{c}(q^{2}), \, \, \, \,
F_{EV}(q^{2})=ZZ_{W}F_{c}^{2}(q^{2}), \, \, \, \, F_{W}(q^{2})=
Z_{W}F_{c}(q^{2}),$$
$$Z_{W}=\frac{1}{2}\{\beta_{V}^{(0)}(Z+N)+\beta_{V}^{(1)}
(Z-N)\},$$
from which $E_{\nu}$ is the neutrino energy, $F_{c}$ is the charge
($F_{c}(0)=1$) contribution form factor of a nucleus with neutrons number
$N=A-Z,$ $\beta_{V}^{(0)}$ and $\beta_{V}^{(1)}$ are the isoscalar and
isovector constants of vector neutral hadronic current.

The terms depending upon $(1+ss')$ and $(1-ss')$ imply that only the
conservation $(s' = s)$ or only the change $(s'=-s)$ of the neutrino helicity
is responsible for the scattering. As seen from (\ref{4}), if the neutrino
has the magnetic moment $F_{2\nu}(q^{2})$ then the left - handed $(s=-1)$ 
neutrino in the reaction (\ref{1}) can be converted into the right - handed 
$(s=+1)$ one, and vice versa \cite{5}. It is clear that interconversions 
\begin{equation}
\nu_{L}\leftrightarrow \nu_{R}, \, \, \, \,
\overline{\nu}_{R}\leftrightarrow \overline{\nu_{L}}
\label{7}
\end{equation}
are carried out in the nucleus field both in reusulting the influence
of the electric charge $F_{1\nu}(q^{2})$ on the neutrino polarization
and in the weak interaction of slow leptons with hadrons. In the latter 
case from (\ref{6}), we are led to the consequence of the Dirac equation 
that the neutrino must change his helicity at the availability 
of a non - zero rest mass \cite{6}.

On the other hand, it is known that in the framework of the $(V-€)$ version of
the theory, between $\mu_{\nu}$ and $m_{\nu}$ there exists a sharp dependence
\cite{7} owing to which, the left - and right - handed neutrinos magnetic
moments answer to the flip of their spins. According to such a point of view,
the processes of interconversions (\ref{7}) originated in the nucleus field
at the expense of form factor $F_{1\nu}(q^{2})$ reflect the availability
of an intimate connection between the mass of the neutrino and
its electric charge.

\end{document}